\documentclass[12pt,preprint]{aastex}
\newcommand{\beq}{\begin{equation}}
\newcommand{\eeq}{\end{equation}}
\newcommand{\ben}{\begin{eqnarray}}
\newcommand{\een}{\end{eqnarray}}

\begin{document}
\title{An Analytic Model for the Axis-Ratio Distribution of 
Dark Matter Halos from the Primordial Gaussian Density Field} 
\author{\sc Jounghun Lee\altaffilmark{1,2}, Y.P. Jing\altaffilmark{3} 
and Yasushi Suto\altaffilmark{4}}

\altaffiltext{1}{School of Physics, Korea Institute for Advanced Study, 
Seoul 207-43, Korea ; jounghun@newton.kias.re.kr}

\altaffiltext{2}{Astronomy Program, School of Earth and Environmental
Sciences, Seoul National University, Seoul 151-742 , Korea}

\altaffiltext{3}{Shanghai Astronomical Observatory; the Partner Group of MPA, 
Nandan Road 80, Shanghai 200030, China; ypjing@shao.ac.cn}

\altaffiltext{4}{Department of Physics, The University of Tokyo, 
 Tokyo 113-0033, Japan ; suto@phys.s.u-tokyo.ac.jp}

\begin{abstract} 
We present an analytic expression for the axis ratio distribution of
triaxial dark matter halos driven from physical principles.  Adopting the 
picture of triaxial collapse based on the Zel'dovich approximation, 
we derive analytically both the minor-to-major and the conditional 
intermediate-to-major axis ratio distributions, and examine how they 
depend on the halo mass, redshift, and cosmology.  Our analytic model 
is tested against the simulation data given by Jing \& Suto in 2002, and 
found to reproduce the conditional intermediate-to-major axis ratio 
distribution successfully and the minor-to-major axis ratio distribution 
approximately.  However, the trends of our analytic axis-ratio distributions 
with mass and redshift are opposite to what is found in N-body simulations. 
This failure of our analytic model puts a limitation on analytic approaches 
based on the Lagrangian theory to the halo ellipticity. 
Given the overall agreement with the simulation 
results, our model provides a new theoretical step toward using the 
axis-ratio distribution of dark halos as a cosmological probe.
We also discuss several possibilities to improve the model.
\end{abstract} 


%%%%%%%%%%%%%%%%%%%%%%%%%%%%%%%%%%%%%%%%%%%%%%%%%%%%%%%%%%%%%%%%%%%%%%%%%%%
\section{INTRODUCTION}

While shapes of dark matter halos have been conventionally modeled as
spherical \citep{nav-etal97,moo-etal99}, optical, X-ray and lensing
observations of galaxy clusters suggest that the shapes of dark halos
are far from spherical \citep{wes89,pli-etal91}.  Recent high-resolution
simulations do indicate that the density profiles of dark halos are
better approximated as triaxial \citep{fre-etal88,war-etal92,jin-sut02,
suw-etal03,kas-evr04,hop-etal04}.

\citet[][JS02 hereafter]{jin-sut02} were able to construct a detailed
empirical model for the triaxial halo for the first time.  Their fitting
model turned out to be quite useful in quantifying the triaxiality
effect on many important observables such as the strong lensing arc
statistics \citep{ogu-etal03,ogu-kee04,dal-etal04}, the halo
mass-temperature relation \citep{yan-etal04}, and the halo figure
rotation \citep{bai-ste04}.

The most fundamental statistics characterizing the halo triaxiality is
the axis-ratio, or equivalently, ellipticity distribution
functions. \citet{hop-etal04} have shown from their N-body simulations
that the axis-ratio distributions depend on the halo environments in
addition to the underlying cosmology.  In order to exploit the halo
axis-ratio distribution as a cosmological probe, therefore, one needs an
analytic model beyond the empirical fitting formulae. This is exactly
what we attempt to propose in this paper.

In fact, the triaxial shape of a dark halo is a generic prediction of
the CDM (Cold Dark Matter) paradigm. \citet{bar-etal86} already derived
an analytic expression of the halo axis-ratio distribution {\it
assuming} that dark halos form at peaks of the primordial Gaussian
density field. They showed theoretically that the CDM dark halos cannot
be spherical as pointed out earlier by \citet{dor70}.  However, recent
numerical results mentioned above have demonstrated that simulated dark
halos are even more elliptical than expected in the previous density
peak approach. This is why we revisit this problem using different
analytical approaches.

To construct a new theoretical model for the halo axis-ratio
distribution, we adopt the cosmic web picture \citep{bon-etal96} which
describes one dimensional filaments in the CDM framework. According to
the picture, the distribution and spatial coherence of initial tidal
fields induces the filamentary pattern of the large scale
structure. Later the filaments bridge between dark matter halos, and the
merging of dark halos occur preferentially along the bridging filaments.
Thus the resulting halos cannot be spherical but naturally become
elongated along the filaments. The halo ellipticities are expected to
increase as the hierarchical merging along the filaments proceeds.
Therefore, the halo axis-ratio can be inferred statistically by
combining an evolution model of non-spherical density perturbations and
the primordial filamentarity of the initial density field.

In order to examine the validity of our analytical model, we compare the
predictions with the simulation results by JS02.  For that purpose, we
consider two specific sets of cosmological parameters that they
adopted. The first model is $\Lambda$CDM model which assumes that
$\Omega_{m}=0.3$, $\Omega_{\Lambda}=0.7$, $\sigma_{8}=0.9$, and
$\Gamma=0.2$, where $\Omega_{m}$ and $\Omega_\Lambda$ denote the matter
density parameter and the dimensionless cosmological constant, and
$\sigma_8$ is the amplitude of the mass fluctuation at $8h^{-1}$Mpc.
Since JS02 use the CDM transfer function of \citet{bar-etal86}
neglecting the baryon contribution, the shape parameter $\Gamma$ simply
equals $\Omega_{m}h$. The second model is SCDM model where
$\Omega_{m}=1$, $\Omega_{\Lambda}=0$, $\sigma_{8}=0.55$, and
$\Gamma=0.5$. JS02 considered dark halos consisting of more than $10^4$
particles. Since a mass of a simulation particle is $2.07\times10^{9}
\Omega_m h^{-1} M_\odot$ ($512^3$ particles in $100h^{-1}$Mpc comoving
cube), it is convenient to define the dimensionless mass of $M_{4}\equiv
M/(2.07\times 10^{13}\Omega_{m}h^{-1}M_{\odot})$.

The rest of this paper is organized as follows. Section 2 describes our
basic assumptions in the analytic modeling.  We lay out mathematical
details of derivation of the axis ratio distributions in \S 3, and
compare the analytic results with their simulation results in \S 4.  
Finally \S 5 is devoted to discussion and conclusions.

\section{BASIC ASSUMPTIONS}

To derive the axis-ratio distribution of dark matter halos, we adopt
three major assumptions. First, the trajectory of a dark matter particle
in the comoving coordinate is well approximated by the Zel'dovich
formula \citep{zel70}. According to the approximation, the key
quantities are the three eigenvalues ($\lambda_{1} \ge \lambda_{2} \ge
\lambda_{3}$) of the deformation tensor (or the tidal shear tensor), 
$d_{ij}$, which is defined as the second derivative of the perturbation 
potential, $\Psi$, at the initial epoch $z_i$:
%%%%%%%%%%%%%%%%%%%%%%%%%%%%%%%%%%%%%%%%%%%%%%%%%%%%%%%%%%%%%%%%%%%%%%%
\begin{equation}
d_{ij} \equiv \partial_{i}\partial_{j}\Psi.  
\end{equation}
%%%%%%%%%%%%%%%%%%%%%%%%%%%%%%%%%%%%%%%%%%%%%%%%%%%%%%%%%%%%%%%%%%%%%%%
The mapping from the Lagrangian to the Eulerian spaces yields an 
expression of the particle density $\rho$ in the Eulerian space ${\bf x}$ 
at redshift $z$: 
%%%%%%%%%%%%%%%%%%%%%%%%%%%%%%%%%%%%%%%%%%%%%%%%%%%%%%%%%%%%%%%%%%%%%%%
\begin{equation}
\label{eqn:density}
\rho({\bf x},z) = \frac{\bar{\rho}(z)}
{[1 - \tilde{D}_{+}(z)\lambda_{1}]
[1 - \tilde{D}_{+}(z)\lambda_{2}]
[1 - \tilde{D}_{+}(z)\lambda_{3}]} ,
\end{equation}
%%%%%%%%%%%%%%%%%%%%%%%%%%%%%%%%%%%%%%%%%%%%%%%%%%%%%%%%%%%%%%%%%%%%%%%
where $\bar{\rho}$ is the background density of the universe, and
$\tilde{D}_{+}(z) \equiv D_{+}(z)/D_{+}(z_{i})$ denotes the linear
growth rate of density fluctuations up to $z$ but normalized to unity at
redshift $z_i$.  Note that $\lambda_{1}$, $\lambda_{2}$, and
$\lambda_{3}$ in equation (\ref{eqn:density}) represent the eigenvalues
of $d_{ij}$ defined at $z_{i}$.  In practice, we use the following
fitting formula \citep{pea99}:
%%%%%%%%%%%%%%%%%%%%%%%%%%%%%%%%%%%%%%%%%%%%%%%%%%%%%%%%%%%%%%%%%%%%%%%
\begin{equation}
\label{eqn:lcdmD}
D_{+}(z) = \frac{5}{2}\Omega_{m}[\Omega_{m}(1+z)^{3}+\Omega_{\Lambda}]^{1/2}
\int_{z}^{\infty}\frac{1+z^{\prime}}
{[\Omega_{m}(1+z^{\prime})^{3} + \Omega_{\Lambda}]^{3/2}}\ dz^{\prime}.
\end{equation}
%%%%%%%%%%%%%%%%%%%%%%%%%%%%%%%%%%%%%%%%%%%%%%%%%%%%%%%%%%%%%%%%%%%%%%%
The {\it linear} density contrast, $\delta$, at $z_i$ is simply written
as
%%%%%%%%%%%%%%%%%%%%%%%%%%%%%%%%%%%%%%%%%%%%%%%%%%%%%%%%%%%%%%%%%%%%%%%
\begin{equation}
\label{eqn:del}
\delta({\bf x},z_i) = \lambda_{1} + \lambda_{2} + \lambda_{3} .
\end{equation}
%%%%%%%%%%%%%%%%%%%%%%%%%%%%%%%%%%%%%%%%%%%%%%%%%%%%%%%%%%%%%%%%%%%%%%%

Second, we assume that a dark matter halo of mass $M$ forms at redshift
$z$ when the corresponding Lagrangian region (at $z_i$) in the linear
density field smoothed over $M$ satisfies the following conditions:
%%%%%%%%%%%%%%%%%%%%%%%%%%%%%%%%%%%%%%%%%%%%%%%%%%%%%%%%%%%%%%%%%%%%%%%
\begin{equation}
\label{eqn:cons}
\delta(z_i) = \delta_{c}(z), \qquad \lambda_{3}(z_i) \ge \lambda_{c}(z),
\end{equation}
%%%%%%%%%%%%%%%%%%%%%%%%%%%%%%%%%%%%%%%%%%%%%%%%%%%%%%%%%%%%%%%%%%%%%%%
where $\delta$ and $\lambda_{3}$ are the linear density contrast and the
smallest eigenvalue of $d_{ij}$ of the smoothed density field, respectively. 
Here $\delta_{c}(z)$ and $\lambda_{c}(z)$ are redshift-dependent threshold 
of $\delta$ and lower limit of $\lambda_{3}$, respectively:
%%%%%%%%%%%%%%%%%%%%%%%%%%%%%%%%%%%%%%%%%%%%%%%%%%%%%%%%%%%%%%%%%%%%%%%
\begin{equation}
\label{eqn:limit}
\delta_{c}(z) = \delta_{c0}\tilde{D}_{+}(0)/\tilde{D}_{+}(z),  \qquad 
\lambda_{c}(z) = \lambda_{c0}\tilde{D}_{+}(0)/\tilde{D}_{+}(z).
\end{equation}
%%%%%%%%%%%%%%%%%%%%%%%%%%%%%%%%%%%%%%%%%%%%%%%%%%%%%%%%%%%%%%%%%%%%%%%
We use the collapse threshold $\delta_{c0}$ computed in the spherical
model.  For SCDM, it is $\delta_{c} \approx 1.686$.  For $\Lambda$CDM,
we use the formula given in Appendix B of \citet{kit-sut96},
which depends weakly on cosmology.

In the spherical approximation, the condition of $\delta(z_i)=
\delta_{c}(z)$ is sufficient for the gravitational collapse at $z$.  In
the non-spherical model based on the Zel'dovich approximation, however,
all regions satisfying $\delta = \delta_{c}$ do not necessarily collapse
into dark halos, since $\lambda_{3}$ can be negative even when $\delta =
\delta_{c}$.  This is why we impose the additional condition
(\ref{eqn:cons}) in our model based on the Zel'dovich approximation.
Nevertheless no reliable modeling is known which determines the value of
$\lambda_{c0}$ from physical principles. While the Zel'dovich
approximation suggests $\lambda_{c0}=1$, those objects are subject to
the first-shell crossing, beyond which the Zel'dovich approximation is
not valid at all. Indeed, \citet{lee-sha98} empirically proposed a value
of $\lambda_{c0}=0.37$ in their mass function theory. They argue that
realistic collapse should proceed along all the three axes almost
simultaneously.  Thus the collapse along the major axis should be
accelerated by the collapse along the other two directions, resulting in
a lower value of $\lambda_{c0}$.  Throughout this paper, we adopt their
value of $\lambda_{c0}=0.37$ unless otherwise stated (see \S 3).

Third, the principal axes of the inertia tensor of a dark matter halo are 
aligned with that of the linear tidal shear tensor of the corresponding 
Lagrangian region. Approximating that the density profile of a dark matter 
halo as a triaxial ellipsoid with three distinct axes, $a$, $b$, and $c$ (we
define $a \le b \le c$), one can say that the inertia shape tensor 
of a dark halo has three distinct eigenvalues, $a,b,c$.   The three 
eigenvalues of the halo inertia tensor, $\{a,b,c\}$, are related to the 
eigenvalues of the tidal shear tensor, $\{\lambda_{1},\lambda_{2}, 
\lambda_{3}\}$ as
%%%%%%%%%%%%%%%%%%%%%%%%%%%%%%%%%%%%%%%%%%%%%%%%
\begin{equation}
\label{eqn:abc}
a \propto \sqrt{1 - \tilde{D}_{+}\lambda_{1}}, \qquad
b \propto \sqrt{1 - \tilde{D}_{+}\lambda_{2}}, \qquad
c \propto \sqrt{1 - \tilde{D}_{+}\lambda_{3}},
\end{equation}
%%%%%%%%%%%%%%%%%%%%%%%%%%%%%%%%%%%%%%%%%%%%%%%%
It may be interesting to compare our definition of the halo axes 
(eq.[\ref{eqn:abc}]) with that of the density peak formalism 
\citep{bar-etal86}. In the density peak formalism, the three eigenvalues 
of the halo inertia tensor are defined as 
\begin{equation}
\label{eqn:den_abc}
a \propto \frac{1}{\sqrt{\zeta_{1}}}, \qquad
b \propto \frac{1}{\sqrt{\zeta_{2}}}, \qquad
c \propto \frac{1}{\sqrt{\zeta_{3}}},
\end{equation}
%%%%%%%%%%%%%%%%%%%%%%%%%%%%%%%%%%%%%%%%%%%%%%%%
where $\zeta_{1}$, $\zeta_{2}$, and $\zeta_{3}$ are the three
eigenvalues of the second derivative of {\it the linear density field},
$\partial_{i}\partial_{j}\delta$.  The comparison of two equations
(\ref{eqn:abc}) and (\ref{eqn:den_abc}) shows that in the density peak
formalism the inertia shape tensor of a dark halo is almost independent
of the tidal shear tensor, $d_{ij}$, while in our formalism which
basically assumes that the ellipticity of a dark halo is induced by the
filamentary cosmic web, it is directly related with $d_{ij}$. In fact,
the strong correlation between the halo inertia and the tidal shear
tensors was demonstrated by recent N-body simulations
\citep{lee-pen00,por-etal02}.
 
The above three assumptions imply that dark matter halos preferentially
form at the over-dense nodes of the filamentary web of the initial
density field where the principal axes of the inertia and the tidal
tensors are aligned with each other. Using these simplified assumptions,
we derive analytically the distribution of the axis-ratios of dark halos
in the following two sections.

\section{THE MINOR-TO-MAJOR AXIS RATIO DISTRIBUTION}
 
We start from the joint probability distribution of the three 
eigenvalues of the tidal shear tensor in the primordial Gaussian density
field \citep{dor70}:  
%%%%%%%%%%%%%%%%%%%%%%%%%%%%%%%%%%%%%%%%%%%%%%%%%%%%%%%%%%%%%%%%%
\begin{equation}
\label{eqn:joint_lam}
p(\lambda_{1},\lambda_{2},\lambda_{3};\sigma_{M}) = 
\frac{3375}{8\sqrt{5}\pi\sigma^6_{M}}
\exp\bigg{(}-\frac{3I_{1}^2}{\sigma^2_{M}}
 + \frac{15I_{2}}{2\sigma^2_{M}}\bigg{)}(\lambda_{1}-\lambda_{2})
(\lambda_{2}-\lambda_{3})(\lambda_{1}-\lambda_{3}) ,
\end{equation}
%%%%%%%%%%%%%%%%%%%%%%%%%%%%%%%%%%%%%%%%%%%%%%%%%%%%%%%%%%%%%%%%%
where 
%%%%%%%%%%%%%%%%%%%%%%%%%%%%%%%%%%%%%%%%%%%%%%%%
\begin{equation}
I_{1} \equiv \lambda_{1}+\lambda_{2}+\lambda_{3},
\qquad 
I_{2} \equiv 
\lambda_{1}\lambda_{2} + \lambda_{2}\lambda_{3} +
\lambda_{3}\lambda_{1},
\end{equation}
%%%%%%%%%%%%%%%%%%%%%%%%%%%%%%%%%%%%%%%%%%%%%%%%%%%%%%%%%%%%%%%%%
We define $\sigma_{M}$ as the rms fluctuation of the linear density
field at $z_i$ smoothed on mass scale $M$:
%%%%%%%%%%%%%%%%%%%%%%%%%%%%%%%%%%%%%%%%%%%%%%%%
\begin{equation}
\label{eqn:sig}
\sigma^{2}_{M}(z_i) = \frac{1}{(2\pi)^{3}}
\int P(k,z_i) W_{TH}^2(kR_M) d^{3} k , 
\end{equation}
%%%%%%%%%%%%%%%%%%%%%%%%%%%%%%%%%%%%%%%%%%%%%%%%
where $P(k,z_i)$ is the linear power spectrum of the density field, and 
$W_{TH}(kR_M)$ is the top-hat filter with 
$R_M \equiv [3M/(4\pi\bar{\rho})]^{1/3}$. 

We change the variables from $\{\lambda_{1},\lambda_{2},\lambda_{3}\}$
to $\{\lambda_{1},\lambda_{2},\delta\}$ using equation (\ref{eqn:del}),
and find the joint probability distribution of
$\lambda_{1}$, $\lambda_{2}$, and $\delta$ from equation
(\ref{eqn:joint_lam}):
%%%%%%%%%%%%%%%%%%%%%%%%%%%%%%%%%%%%%%%%%%%%%%%%
\begin{eqnarray}
p(\lambda_{1},\lambda_{2},\delta ; \sigma_{M}) &=& 
\frac{3375}{8\sqrt{5}\pi\sigma^{6}_{M}}
\exp\left[-\frac{3\delta^{2}}{\sigma^{2}_{M}} + 
\frac{15\delta(\lambda_{1}+\lambda_{2})}{2\sigma^{2}_{M}} -
\frac{15(\lambda^{2}_{1}+\lambda_{1}\lambda_{2}+\lambda^{2}_{2})}
{2\sigma^{2}_{M}}\right] \nonumber \\
&&\times(2\lambda_{1}+\lambda_{2}-\delta)(\lambda_{1}-\lambda_{2})
(\lambda_{1}+2\lambda_{2}-\delta).
\label{eqn:lam_del}
\end{eqnarray}
%%%%%%%%%%%%%%%%%%%%%%%%%%%%%%%%%%%%%%%%%%%%%%%%
Applying the Bayes theorem and the Gaussian distribution of the linear
density
%%%%%%%%%%%%%%%%%%%%%%%%%%%%%%%%%%%%%%%%%%%%%%%%
\begin{equation}
p(\delta;\sigma_{M}) = \frac{1}{\sqrt{2\pi}\sigma_{M}}
\exp\left(-\frac{\delta^{2}}{\sigma^{2}_{M}}\right),
\end{equation}
%%%%%%%%%%%%%%%%%%%%%%%%%%%%%%%%%%%%%%%%%%%%%%%%
the conditional probability distribution of $\lambda_{1}$ and 
$\lambda_{2}$ at $\delta = \delta_{c}$ is written as 
%%%%%%%%%%%%%%%%%%%%%%%%%%%%%%%%%%%%%%%%%%%%%%%%
\begin{eqnarray} 
p(\lambda_{1},\lambda_{2} | \delta = \delta_{c} ; \sigma_{M}) &=& 
\frac{p(\lambda_{1},\lambda_{2},\delta = \delta_{c} ; \sigma_{M})
d\delta}{p(\delta = \delta_{c}; \sigma_{M})d\delta} \nonumber \\
&=&\frac{3375\sqrt{2}}{\sqrt{10\pi}\sigma^{5}_{M}}
\exp\left[-\frac{5\delta^{2}_{c}}{2\sigma^{2}_{M}} + 
\frac{15\delta_{c}(\lambda_{1}+\lambda_{2})}{2\sigma^{2}_{M}} -
\frac{15(\lambda^{2}_{1}+\lambda_{1}\lambda_{2}+\lambda^{2}_{2})}
{2\sigma^{2}_{M}}\right] \nonumber \\
&&\times(2\lambda_{1}+\lambda_{2}-\delta_{c})(\lambda_{1}-\lambda_{2})
(\lambda_{1}+2\lambda_{2}-\delta_{c}).
\label{eqn:con_lamdel}
\end{eqnarray}
%%%%%%%%%%%%%%%%%%%%%%%%%%%%%%%%%%%%%%%%%%%%%%%%
 
We define two {\it real} variables, $\mu_{1}$ and $\mu_{2}$, as the axis
ratios of a triaxial halo:
%%%%%%%%%%%%%%%%%%%%%%%%%%%%%%%%%%%%%%%%%%%%%%%%%%%%%%%%%%%%%%%%
\begin{equation}
\mu_{1} \equiv \frac{b}{c},  \qquad \mu_{2} \equiv \frac{a}{c}
\end{equation}
%%%%%%%%%%%%%%%%%%%%%%%%%%%%%%%%%%%%%%%%%%%%%%%%%%%%%%%%%%%%%%%%
($\mu_2 \le \mu_1 \le 1$).  According to the third assumption in \S 2,
$\mu_{1}$ and $\mu_{2}$ are written in terms of $\lambda_{1}$,
$\lambda_{2}$, and $\delta$ as
%%%%%%%%%%%%%%%%%%%%%%%%%%%%%%%%%%%%%%%%%%%%%%%%%%%%%%%%%%%%%%%%
\begin{equation}
\label{eqn:mus}
\mu_{1} = \sqrt{\frac{1 - \tilde{D}_{+}\lambda_{2}}
{1 - \tilde{D}_{+}(\delta_{c}-\lambda_{1}-\lambda_{2})}}, \qquad
\mu_{2} = \sqrt{\frac{1 - \tilde{D}_{+}\lambda_{1}}
{1 - \tilde{D}_{+}(\delta_{c}-\lambda_{1}-\lambda_{2})}}, 
\end{equation}
%%%%%%%%%%%%%%%%%%%%%%%%%%%%%%%%%%%%%%%%%%%%%%%%%%%%%%%%%%%%%%%%
with the following constraints:
%%%%%%%%%%%%%%%%%%%%%%%%%%%%%%%%%%%%%%%%%%%%%%%%%%%%%%%%%%%%%%%%
\begin{equation}
\delta_{c}-\lambda_{1}-\lambda_{2} \ge \lambda_{c}(z), \qquad 
\lambda_{1} \le \frac{1}{\tilde{D}_{+}(z)} .
\end{equation}
%%%%%%%%%%%%%%%%%%%%%%%%%%%%%%%%%%%%%%%%%%%%%%%%%%%%%%%%%%%%%%%%
The first constraint $\lambda_{3} \ge \lambda_{c}(z)$ guarantees the
collapse along all three axes, in accordance with the second assumption
in $\S 2$.  The second constraint $\lambda_{1} \le 1/\tilde{D}_{+}(z)$
guarantees that $\mu_{1}$ and $\mu_{2}$ are all real. Note that if
$\lambda_{1} \le 1/\tilde{D}_{+}(z)$, then it automatically implies
$\lambda_{3} \le \lambda_{2} \le 1/\tilde{D}_{+}(z)$. According to equation
(\ref{eqn:con_lamdel}), however, $\lambda_{1}$ has a non-zero
probability of $\lambda_{1} > 1/\tilde{D}_{+}(z)$.  We simply do not
consider the parameter region of $\lambda_{1} \ge 1/\tilde{D}_{+}(z)$ since
they correspond to the break-down of the Zel'dovich approximation in the
non-linear regime after the first-shell crossing ($\lambda_{1} =
1/\tilde{D}_{+}(z)$).

Now, we write the probability density that a dark matter halo of mass $M$ 
formed at redshift $z_{f}$ has a intermediate-to-major axis ratio of $b/c$ 
and the minor-to-major axis ratio of $a/c$ as 
%%%%%%%%%%%%%%%%%%%%%%%%%%%%%%%%%%%%%%%%%%%%%%%%%%%%%%%%%%%%%%%%
\begin{eqnarray}
\label{eqn:ratio_dis}
p(b/c, a/c ; M, z_{f}) &\equiv& 
p(\mu_{1},\mu_{2}| \delta = \delta_{c} ; \sigma_{M};z_{f}) \cr
&=& A\, p(\lambda_{1},\lambda_{2} | \delta=\delta_c ; \sigma_{M}; z_{f})
\Theta\!\left(\frac{1}{\tilde{D}_f}-\lambda_{1}\right) \cr
&& \times \Theta[\delta_{c}-\lambda_{c}-(\lambda_{1}+\lambda_{2})]
\left|\frac{(\partial\lambda_{1}\partial\lambda_{2})}
{(\partial\mu_{1}\partial\mu_{2})}\right|,
\end{eqnarray}
%%%%%%%%%%%%%%%%%%%%%%%%%%%%%%%%%%%%%%%%%%%%%%%%%%%%%%%%%%%%%%%%
where we solve equation (\ref{eqn:mus}) for
$\lambda_{1}$, $\lambda_{2}$ and $\lambda_{3}$ as 
%%%%%%%%%%%%%%%%%%%%%%%%%%%%%%%%%%%%%%%%%%%%%%%%%%%%%%%%%%%%%%%%
\begin{eqnarray}
\label{eqn:lamu1}
\lambda_{1} &=& \frac{1 + (\tilde{D}_f\delta_{c}- 2)\mu^{2}_{2} + 
\mu^{2}_{1}}
{\tilde{D}_f(\mu^{2}_{1} + \mu^{2}_{2} + 1)},\\
\label{eqn:lamu2} 
\lambda_{2} &=& \frac{1 + (\tilde{D}_f\delta_{c}- 2)\mu^{2}_{1} + 
\mu^{2}_{2}}{\tilde{D}_f
(\mu^{2}_{1} + \mu^{2}_{2} + 1)},\\
\label{eqn:lamu3} 
\lambda_{3} &=& \frac{\tilde{D}_f\delta_{c}- 2 +\mu^{2}_{1} + 
\mu^{2}_{2}}{\tilde{D}_f
(\mu^{2}_{1} + \mu^{2}_{2} + 1)}.
\end{eqnarray}
%%%%%%%%%%%%%%%%%%%%%%%%%%%%%%%%%%%%%%%%%%%%%%%%%%%%%%%%%%%%%%%%
In the above expression, $\delta_{c}$ and $\tilde{D}_f$ depend on
the formation redshift, $z_{f}$: $\delta_{c}(z_{f})$ and
$\tilde{D}_f = \tilde{D}_{+}(z_{f})$, and $\Theta$ is the Heaviside
step function. The normalization constant $A$ satisfies
%%%%%%%%%%%%%%%%%%%%%%%%%%%%%%%%%%%%%%%%%%%%%%%%%%%%%%%%%%%%%%%%
\begin{equation}
A\int p(\lambda_{1},\lambda_{2} | \delta=\delta_c ; \sigma_{M}, z_{f})
\Theta\!\left(\frac{1}{\tilde{D}_f}-\lambda_{1}\right)
\Theta[\delta_{c}-\lambda_{c}-(\lambda_{1}+\lambda_{2})]
d\mu_{1}d\mu_{2} = 1.
\label{eqn:norm_const}
\end{equation}
%%%%%%%%%%%%%%%%%%%%%%%%%%%%%%%%%%%%%%%%%%%%%%%%%%%%%%%%%%%%%%%%
Finally we denote by $\left|(\partial\lambda_{1}\partial\lambda_{2})/
(\partial\mu_{1}\partial\mu_{2})\right|$ the Jacobian of the
transformation from $\{\lambda_{1},\lambda_{2}\}$ to
$\{\mu_{1},\mu_{2}\}$, which we find from equations
(\ref{eqn:lamu1}) and (\ref{eqn:lamu2}):
%%%%%%%%%%%%%%%%%%%%%%%%%%%%%%%%%%%%%%%%%%%%%%%%%%%%%%%%%%%%%%%%
\begin{equation}
\label{eqn:jac}
\left|\frac{(\partial\lambda_{1}\partial\lambda_{2})}
{(\partial\mu_{1}\partial\mu_{2})} \right| = 
\frac{4(\tilde{D}_f\delta_{c} - 3)^2\mu_{1}\mu_{2}}
{\tilde{D}_f^2(\mu^{2}_{1}+\mu^{2}_{2}+1)^{3}}.
\end{equation}
%%%%%%%%%%%%%%%%%%%%%%%%%%%%%%%%%%%%%%%%%%%%%%%%%%%%%%%%%%%%%%%%

Integrating equation (\ref{eqn:ratio_dis}) over $b/c$, we find the
probability density that a dark halo of mass $M$ formed at redshift
$z_{f}$ has a minor-to-major axis ratio of $a/c$:
%%%%%%%%%%%%%%%%%%%%%%%%%%%%%%%%%%%%%%%%%%%%%%%%%%%%%%%%%%%%%%%%
\begin{eqnarray}
\label{eqn:capro}
p(a/c ; M, z_{f}) &=& \int_{a/c}^{1} p(b/c, a/c ; M, z_{f}) d(b/c)
\cr
&=& \int_{\mu_2}^{1}
p[\mu_{1},\mu_{2}| \delta = \delta_{c}(z_{f}); \sigma_{M}]d\mu_{1}
\cr
&=& \int_{\mu_2}^{1}
A\ p[\lambda_{1},\lambda_{2}|\delta=\delta_c(z_{f});\sigma_{M}]
\Theta\!\left(\frac{1}{\tilde{D}_f}-\lambda_{1}\right)
\cr
&& \times \Theta[\delta_{c}-\lambda_{c}-(\lambda_{1}+\lambda_{2})]
\left|\frac{(\partial\lambda_{1}\partial\lambda_{2})}
{(\partial\mu_{1}\partial\mu_{2})} \right|d\mu_{1} .
\end{eqnarray}
%%%%%%%%%%%%%%%%%%%%%%%%%%%%%%%%%%%%%%%%%%%%%%%%%%%%%%%%%%%%%%%%

Equation (\ref{eqn:capro}) is the axis ratio distribution at the
formation epoch, $z_{f}$. The axis-ratio distribution at the {\it
observation epoch}, $z$, can be readily found as
%%%%%%%%%%%%%%%%%%%%%%%%%%%%%%%%%%%%%%%%%%%%%%%%%%%%%%%%%%%%%%%%
\begin{equation}
\label{eqn:capro_z}
p(a/c ; M ; z) = \int_{z}^{\infty}dz_{f}\ 
 \frac{\partial p_{f}(z_{f};2M,z)}{\partial z_{f}}\ p(a/c ; 2M ; z_{f}),
\end{equation}
%%%%%%%%%%%%%%%%%%%%%%%%%%%%%%%%%%%%%%%%%%%%%%%%%%%%%%%%%%%%%%%%
where the formation epoch distribution, $\partial p_{f}/\partial z_{f}$,
represents the probability that a halo of mass $2M$ that exists at $z$
had a mass greater than $M$ for the first time at $z_{f}$.
Since the formation epoch distribution in the current non-spherical model 
is almost impossible to work out analytically, we use the spherical
counterpart instead. In practice we use the fitting formula by
\citet{kit-sut96} for the analytic expression derived by
\citet{lac-col94}. 

Let us emphasize the difference between equations (\ref{eqn:capro}) and
(\ref{eqn:capro_z}): while the former gives the probability density that
a dark halo of mass $M$ has a minor-to-major axis ratio $a/c$ at its
formation redshift $z_f$, the latter is the counterpart evaluated
at the observation redshift $z (<z_f)$.  Of course, in numerical
simulations and observations, the latter, not the former, is the
relevant observable.  In what follows, therefore, we mainly consider the
latter evaluated at the {\it observation} epoch, $z$.

\begin{figure}
\begin{center}
%\leavevmode\epsfxsize=11.0cm \epsfbox{f1.eps}
\plotone{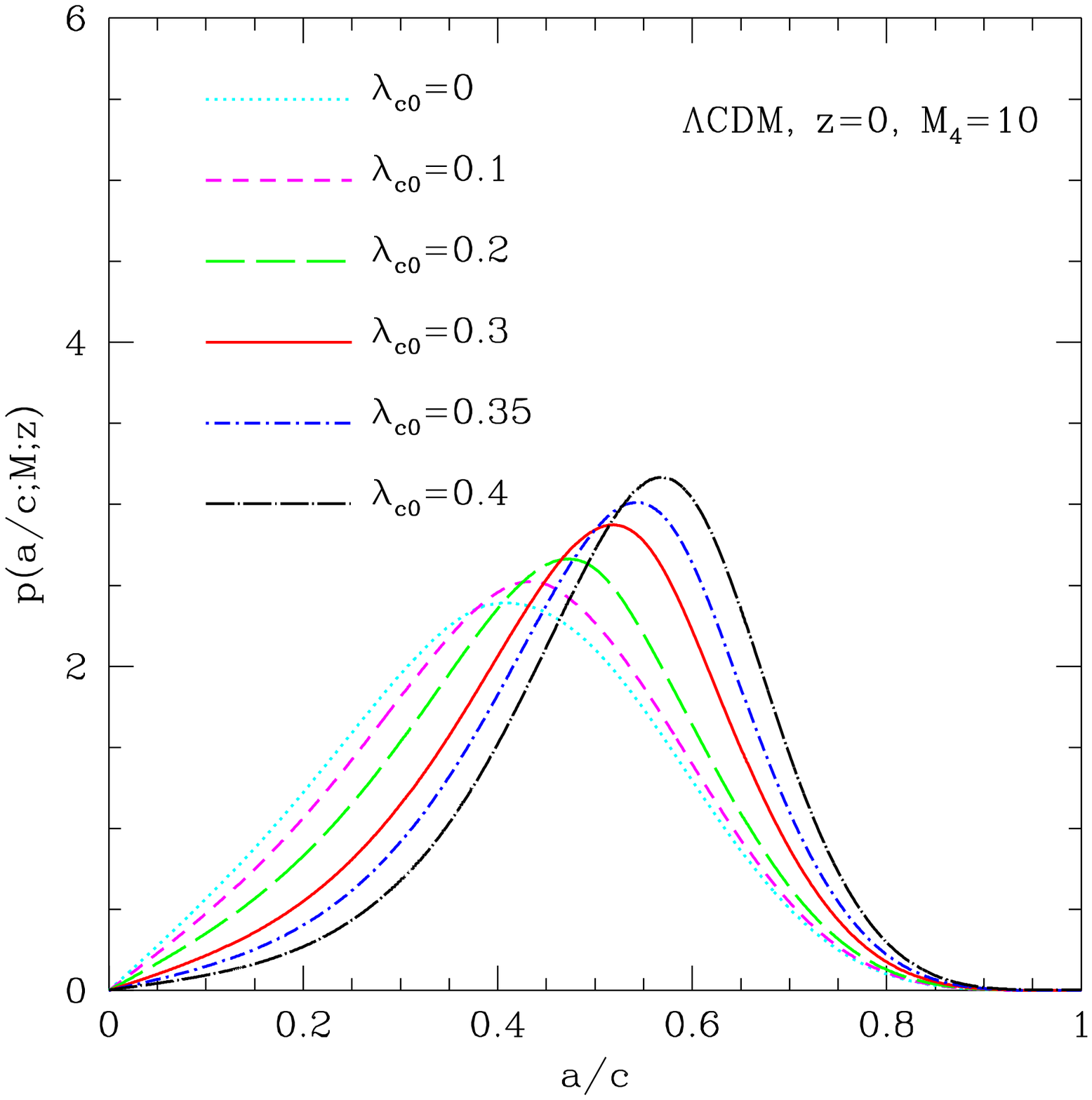} \caption{Probability density distribution of the axis
ratio $a/c$ of halos for the five different cases of the short-axis
cut-off in the $\Lambda$ CDM model; $\lambda_{c0}=0$, $0.1$, $0.2$,
$0.3$, $0.35$ and $0.4$ (dotted, dashed, long-dashed, solid, dot-dashed,
and dot-long dashed lines) respectively. Here, the halo mass and redshift
are set to be $M_{4} \equiv M/(2.07\times 10^{13}\Omega_{m}h^{-1}M_{\odot}) 
= 10$ and $z=0$, respectively. \label{fig:lamc}}
\end{center}
\end{figure}

Figure \ref{fig:lamc} plots the $\lambda_{c0}$-dependence of $p(a/c ; M,
z)$ in the $\Lambda$CDM model: $\lambda_{c0} = 0$, $0.1$, $0.2$, $0.3$,
$0.35$ and $0.4$ (dotted, dashed, long-dashed, solid, dot-dashed, and
dot-long-dashed lines, respectively). For this plot, we choose the halo
mass $M_{4} = 10$ at $z=0$ for definiteness.  The axis-ratio
distribution of dark halos shifts toward the high axis-ratio side (more
spherical) as $\lambda_{c0}$ increases. This is theoretically
understandable since higher values of $\lambda_{c0}$ correspond to those
Lagrangian regions whose major axis lengths are closer to the other two
(see eq.[\ref{eqn:abc}]).  Given such, we adopt $\lambda_{c0}=0.37$ in
the analysis below, following \citet{lee-sha98}

\begin{figure}
\begin{center}
%\leavevmode\epsfxsize=11.0cm \epsfbox{f2.eps}
\plotone{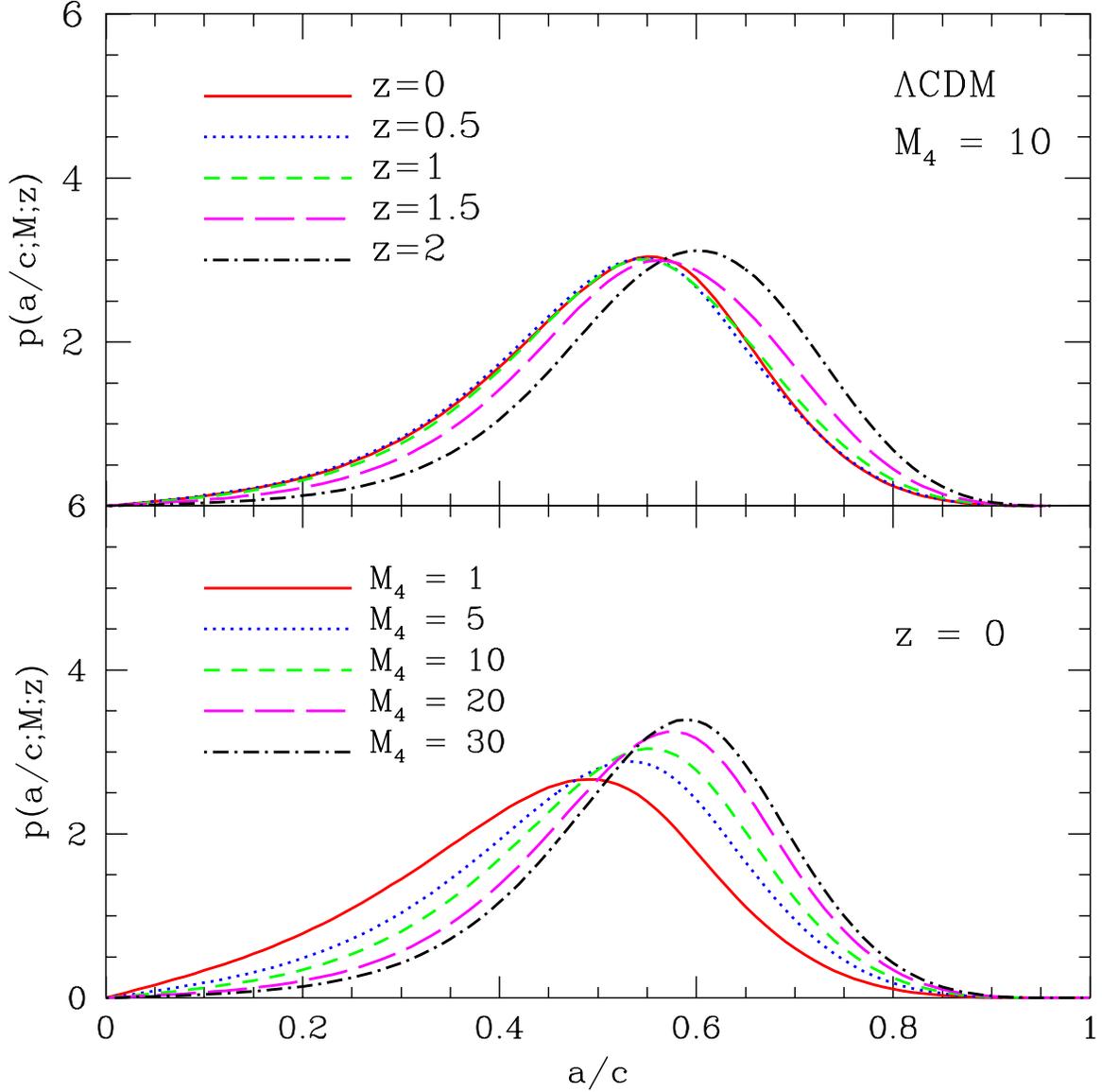}
\caption{Behavior of the probability density distributions of the axis ratio 
$a/c$ of halos with the change of observation epochs $z_{f}$ and the halo mass 
$M$ in the $\Lambda$ CDM model. 
{\it Top}: Case in which  $z = 0$, $0.5$, $1$, $1.5$ and $2$ 
(solid, dotted, dashed, long-dashed, and dot-dashed, respectively).
{\it Bottom}: Case in which $M_{4} = 1$, $5$, $10$, $20$ and $30$ 
(solid, dotted, dashed, long-dashed, and dot-dashed,respectively), 
where $M_{4} \equiv M/(2.07\times 10^{13}\Omega_{m}h^{-1}M_{\odot})$.
\label{fig:redm}}
\end{center}
\end{figure}

Figure \ref{fig:redm} shows how $p(a/c;M,z)$ depends on the halo mass
and the redshift for the case of $\Lambda$CDM.  The upper panel shows
the $z$-dependence of $p(a/c ; M, z)$ for a fixed mass scale of
$M_{4}=10$ at redshift $z = 0$, $0.5$, $1$, $1.5$ and $2$ (solid,
dotted, dashed, long-dashed, and dot-dashed) respectively. While the
lower panel plots the mass-dependence of $p(a/c ; M, z)$ at $z=0$ for
$M_{4} = 1,5,10,20$ and $30$ (solid, dotted, dashed, long-dashed, and
dot-dashed) respectively.  As can be seen, the analytic distribution
$p(a/c;M,z)$ depends on the halo mass and redshift consistently: the
distribution moves toward the high axis-ratio section as the halo mass
and the redshift increase.

In other words, our analytic model predicts that the more massive a dark 
halos is, the less elliptical it is, and that a dark halo of given mass 
is less elliptical at earlier epochs. These two trends are consistent 
with the theoretical work of \citet{ber94}. By applying the perturbation 
theory to a primordial Gaussian density field, he proved theoretically 
that the larger halos formed at higher density peaks are rounder. 
However, these trends are opposite to what is found through accurate 
calculations of N-body simulations: It was found in simulations that the 
more massive halos are more elliptical, and that the halos of given mass 
observed at earlier epochs are more elliptical 
\citep{bul02,spr-etal04,jin-sut02,hop-etal04}. 
This failure of our analytic model implies that the dependences of the 
axis-ratio distributions on mass and redshift are not fully determined 
by simply applying the Zel'dovich approximation.

\begin{figure}
\begin{center}
%\leavevmode\epsfxsize=11.0cm \epsfbox{f3.eps}
\plotone{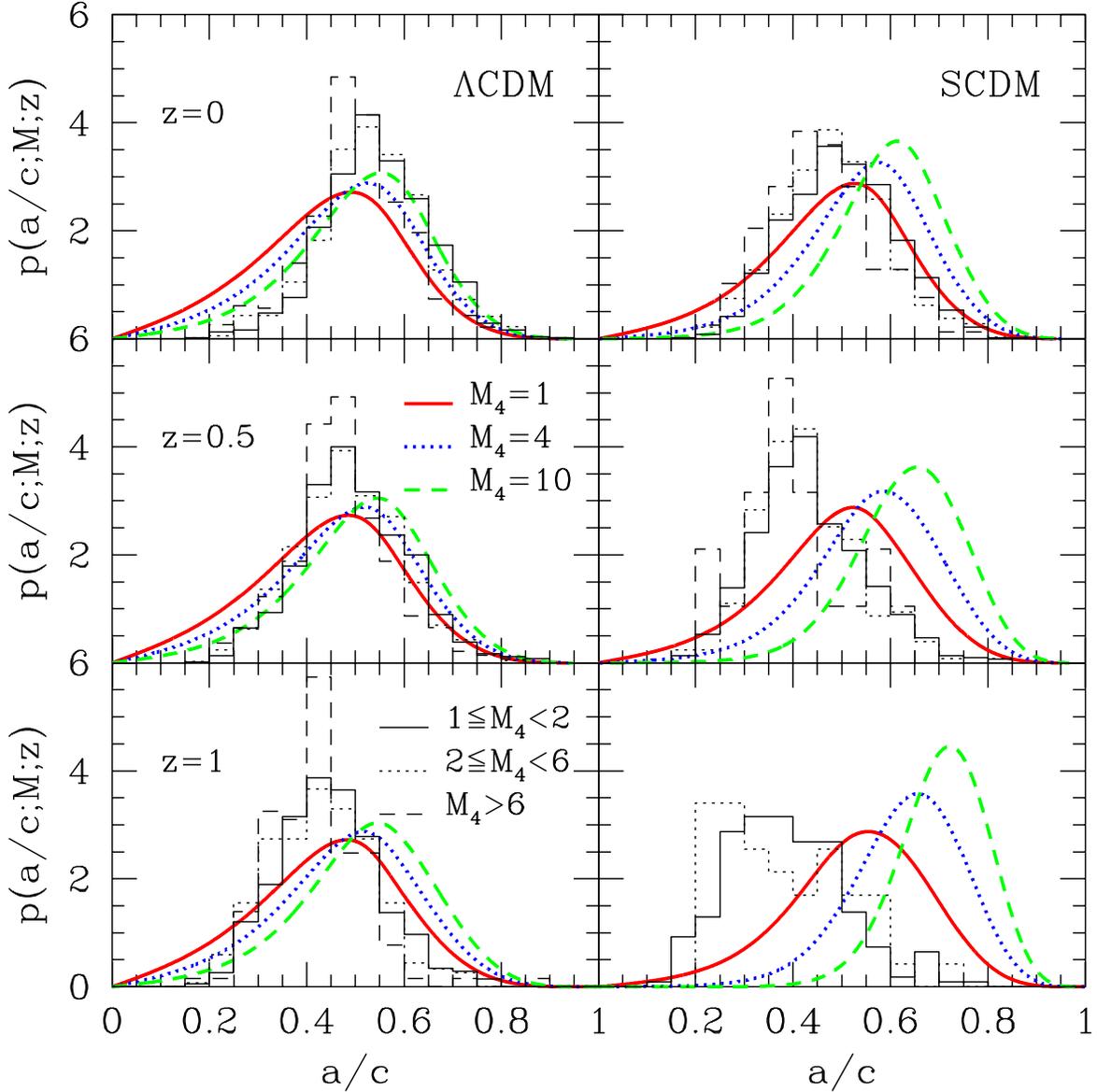}
\caption{Probability density distributions of the axis ratio $a/c$ of halos 
in the $\Lambda$ CDM (left) and SCDM (right) models at three different 
observation  epochs; $z=0$, $0.5$,and $1$ (top, middle, and bottom panels, 
respectively), with the choice of $\lambda_{c0}=0.37$.  In each panel, 
the histograms represent the numerical results \citep{jin-sut02} for 
those halos of mass; $1\le M_{4} < 2$, $2\le M_{4} < 6$, and $M_{4} \ge 6$ 
(thin solid, dotted, and dashed lines,  respectively.) where 
$M_{4}\equiv M/(2.07\times 10^{13}\Omega_{m}h^{-1}M_{\odot})$, while 
the curves represent the analytic results (see eq.[\ref{eqn:capro}]) 
for those halos of mass; $M_{4} = 1$, $M_{4} = 4$, and $M_{4} =10$ 
(thick solid, dotted, and dashed lines), respectively. 
\label{fig:pro_ac}}
\end{center}
\end{figure}

Figure \ref{fig:pro_ac} shows our analytic prediction
(eq.[\ref{eqn:capro_z}]) at $z = 0$, $0.5$, and $1$ (top, middle, and
bottom, respectively) for three mass scales $M_{4}=1,4,10$ (solid,
dotted, and dashed, respectively). They should be compared with the
numerical results (histograms) from JS02 for the $\Lambda$CDM
(left panels) and the SCDM (right panels).  Clearly the analytic
predictions agree with the numerical results reasonably well; they
reproduce well shapes and characteristic behaviors (especially for
$\Lambda$CDM) such as the peak positions, the dispersions, and the
decrease of the mean axis-ratios with the increase of redshift.  On
the other hand, we notice that the numerical histograms slightly move
toward the low axis ratio section as the halo mass increases, which
disagrees with the analytical predictions. We discuss on this
disagreement between the analytical and the numerical results in $\S 5$.

\section{THE CONDITIONAL INTERMEDIATE-TO-MAJOR AXIS RATIO DISTRIBUTION}

The probability density, $p(b/c;M;z)$, that a dark halo of mass $M$ is 
observed at redshift $z$ to have an intermediate-to-major axis 
ratio of $b/c$ can be also computed in a similar manner. However, to
investigate the overall triaxiality of a dark halo, what is more
relevant is the {\it conditional} probability density distribution,
$p(b/c|a/c ;M;z)$, that a dark halo of mass $M$ is observed at $z$
to have an intermediate-to-major axis ratio $b/c$ provided it has a
minor-to-major axis ratio $a/c$.  In principle, one can find this
conditional probability density from the Bayes theorem:
%%%%%%%%%%%%%%%%%%%%%%%%%%%%%%%%%%%%%%%%%%%%%%%%%%%%%%%%%%%%%%%%
\begin{equation}
\label{eqn:bapro_zf}
p(b/c | a/c ; M ; z) = \frac{p(b/c, a/c ; M ; z)d(a/c)}
{p(a/c ; M ; z)d(a/c)}.
\end{equation}
%%%%%%%%%%%%%%%%%%%%%%%%%%%%%%%%%%%%%%%%%%%%%%%%%%%%%%%%%%%%%%%%
Although it is possible to derive $p(b/c | a/c ; M ; z)$ analytically as
well, it is not easy to construct the statistical sample either from the
current simulations or from observations; there would be only a few dark
halos of mass $M$ at redshift $z$ with the fixed minor-to-major $a/c$.

To overcome the poor number statistics, JS02 combined all the halo mass,
i.e., they computed $p(b/c | a/c ; z)$ instead of $p(b/c | a/c ; M ;
z)$.  For a direct comparison with their result, we compute $p(b/c | a/c
; z)$ according to
%%%%%%%%%%%%%%%%%%%%%%%%%%%%%%%%%%%%%%%%%%%%%%%%%%%%%%%%%%%%%%%%
\begin{equation}
\label{eqn:bapro}
p(b/c | a/c ; z) = \int_{S_{M}}dM\int_{z}^{\infty}dz_{f}\ n(2M;z)
 \frac{\partial p_{f}(z_{f};2M,z)}{\partial z_{f}}\ p(b/c | a/c ; 2M ; z_{f}), 
\end{equation}
%%%%%%%%%%%%%%%%%%%%%%%%%%%%%%%%%%%%%%%%%%%%%%%%%%%%%%%%%%%%%%%%
where $p(b/c | a/c ; 2M ; z_{f})$ is given as equation
(\ref{eqn:ratio_dis}), $n(M;z)$ represents the number density of dark
halos of mass $M$ that exist at $z$, and $S_{M}$ represents the whole
mass range to be considered.  In practice, we use the fitting formula
given by \citet{she-tor99} for $n(M;z)$.

\begin{figure}
\begin{center}
%\leavevmode\epsfxsize=11.0cm \epsfbox{f4.eps}
\plotone{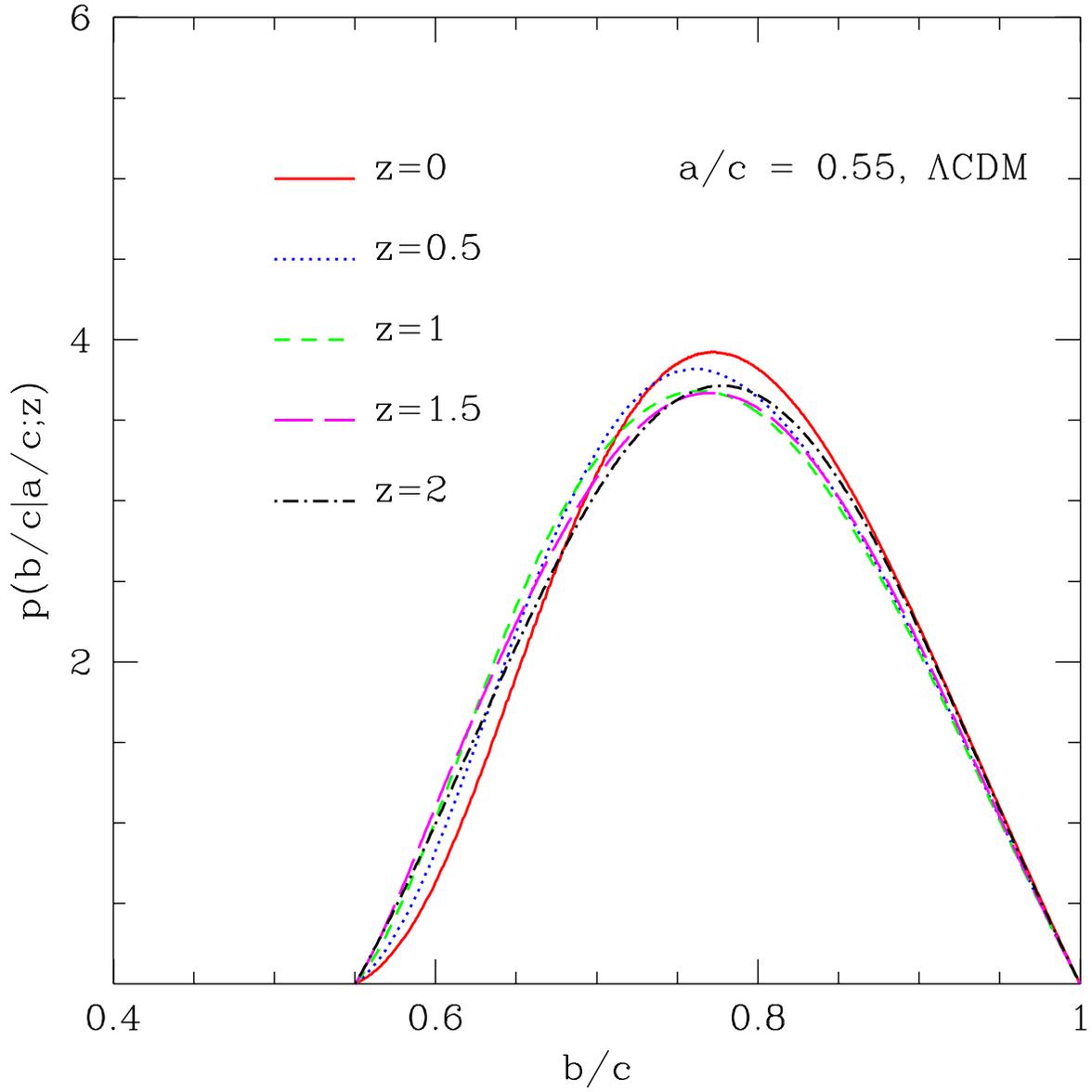}
\caption{Behavior of the conditional probability density distributions of 
the axis ratio $b/c$ at $z = 0$, $0.5$, $1$, $1.5$ 
and $2$ (solid, dotted, dashed, long-dashed, and dot-dashed), respectively. 
\label{fig:beh}}
\end{center}
\end{figure}

Figure \ref{fig:beh} illustrates the redshift-dependence of $p(b/c |
a/c; z)$ in the $\Lambda$CDM model at $z=0$, $0.5$, $1$, $1.5$ and $2$
(solid, dotted, dashed, long-dashed, and dot-dashed) respectively. The
minor-to-major axis ratio is fixed to be $a/c = 0.55$.  As one can see,
the conditional distribution $p(b/c | a/c; z)$ is insensitive to the
redshift, which is also consistent with the finding of JS02.

\begin{figure}
\begin{center}
%\leavevmode\epsfxsize=11.0cm \epsfbox{f5.eps}
\plotone{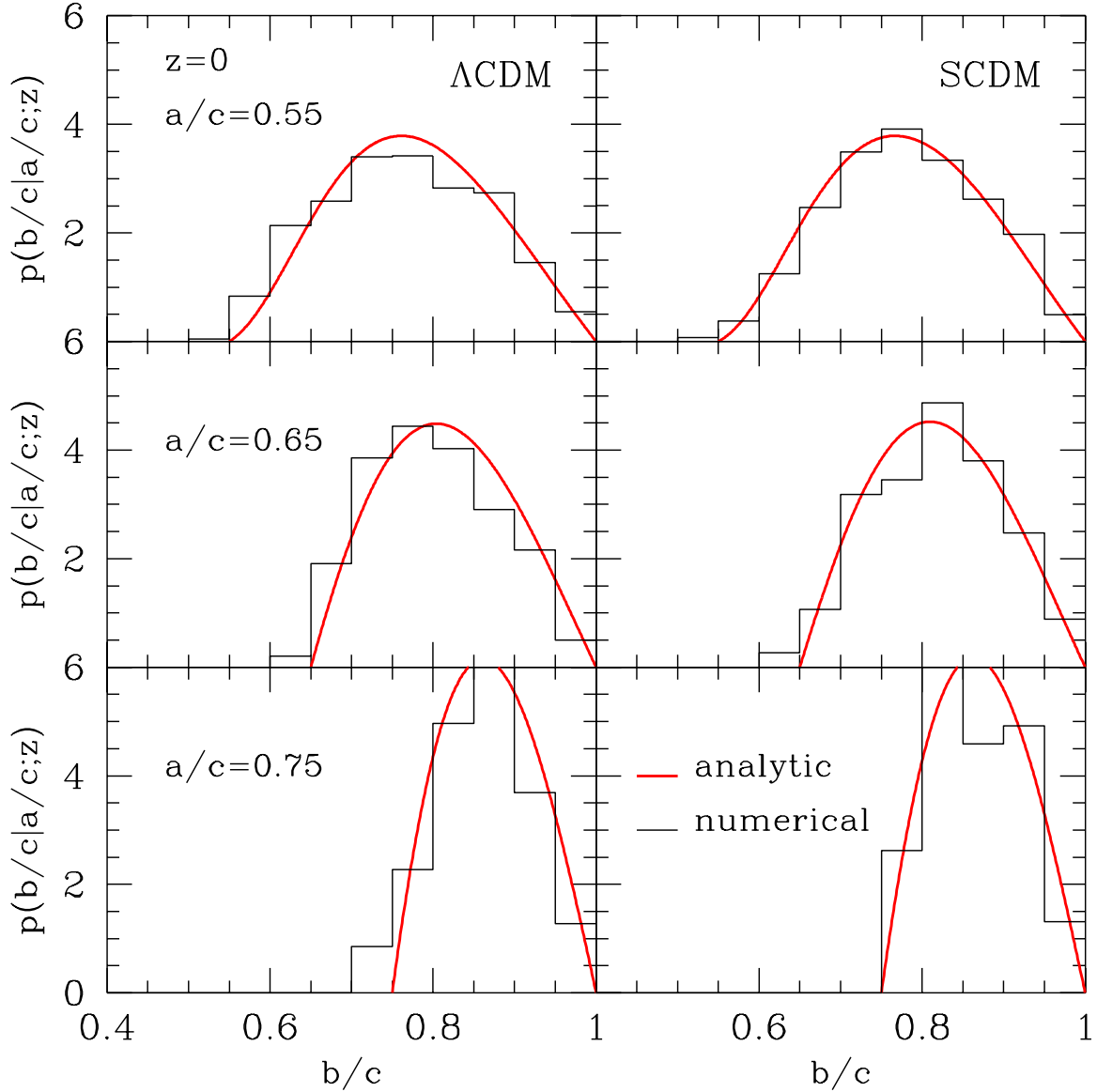}
\caption{Conditional probability density distributions of the axis ratio 
$a/c$ of halos in the $\Lambda$ CDM (left) and SCDM (right) models 
provided that the minor-to-major axis ratio has a certain value: 
$a/c = 0.55$, $0.65$ and $a/c = 0.75$ (top, middle, and bottom panels),
respectively at $z=0$.  In each panel, the histogram and the curve 
represent the numerical \citep{jin-sut02} and the analytic results, 
respectively. 
\label{fig:con_pro}}
\end{center}
\end{figure}

Figure \ref{fig:con_pro} compares the analytic predictions (curves) with
the numerical findings (histograms) for $a/c = 0.55$, $0.65$, and $0.75$
(top, middle, and bottom, respectively) at $z=0$ in $\Lambda$CDM (left
panels) and SCDM (right panels) models.  The histograms are computed by
averaging over $0.5 \le a/c < 0.6$, $0.6 \le a/c < 0.7$, and $0.7 \le
a/c < 0.8$ (top, middle, and bottom, respectively).  The agreement
between analytic and numerical results is quite satisfactory.

\section{DISCUSSIONS AND CONCLUSIONS}

We have derived an analytic expression for the axis-ratio distribution
of triaxial dark matter halos for the first time. In constructing the
analytic model, we adopted the cosmic web picture in which the
ellipticities of dark matter halos are induced by the coherent tidal
fields in the initial density fluctuations, and employed the
Zel'dovich-type collapse condition as a diagnostics.  Our analytic model
is successful in reproducing the basic behaviors on a qualitative level
found from the previous numerical simulations (JS02). This ensures our 
basic picture that the origin of the halo triaxiality is the filamentarity 
of the initial density field.  Nevertheless we found a discrepancy with 
the numerical results on a quantitative level; in particular the predicted 
dependences of the halo triaxiality on mass and redshift seem inconsistent 
with the numerical results.

In order for the future improvements, let us critically discuss possible
caveats in our analytic approaches: First of all, we use a cooked-up
collapse condition of $\delta=\delta_{c}$ and $\lambda_{3} \ge\lambda_{c}$. 
What we really mean to have is a practical and simple collapse condition 
for the formation of a triaxial dark halo in the nodes of filamentary web 
of the density field by combining the density peak formalism and the 
Zel'dovich approximation. This collapse condition is theoretically unjustified.

How to identify a triaxial halo and where to locate the corresponding 
Lagrangian site in the linear density field is a touchy issue in the 
non-spherical dynamical model.  Although \citet{bon-mye96} proposed the peak 
patch picture as a complete non-spherical model, their formalism is too 
complicated to follow analytically in practice. Besides, as discussed in 
\citet{sut02}, there is still no unanimous agreement among theory, 
observations, and simulations about how to define a gravitationally bound 
object in practice. The disagreement among the three become more serious 
when a dark halo is to be described as triaxial. 
Hence, to find a theoretically justified criterion for the formation of a 
triaxial dark halo and to derive the axis-ratio distribution more rigorously 
with the criterion, it will be necessary to address this difficult issue 
first, which is beyond the scope of this paper. 

As \citet{ber94} proved theoretically that a rare event in the linear 
density field is inclined to be quite spherical. In other words, a 
linear over-dense region of high-mass must be more spherical than a 
low-mass over-dense region. This is the case that our analytic 
distribution predicts. The difference in the trends with mass and redshift 
between our model and the simulation results may be caused by the fact 
that in reality the shape of a dark halo must be affected by subsequent 
nonlinear clustering process. We argue here that the nonlinear 
merging event must play a key role in increasing the halo ellipticity. 
Many N-body studies demonstrated that the merging event occurs
anisotropically along filaments \citep[e.g.,][]{wes-etal91,van-van93,dub98,
fal-etal02}. This anisotropic merging event tends to make the shapes of 
dark halos elongated by aligning their substructures along with the 
orientation of their major axes \citep[e.g.,][]{kne-etal04}. Moreover, 
it was also shown that the merging process is more rapid for the case 
of higher mass halos \citep{zha-etal03}.  The tendency of being 
more spherical in the higher-mass section of the linear density field 
\citep{ber94} is likely to be compensated by the anisotropic merging 
effect. Therefore, the overall dependence of the axis-ratio distribution 
on the halo mass and redshift may be weakened by this compensating effect. 
Our future work is in the direction of incorporating semi-analytically 
the anisotropic merging process into our analytic model.

In passing, it is interesting to note that the anisotropic merging of
dark halos along filaments has an implication about the mass function of
dark halos.  In the standard mass function theory based on the
Press-Schechter theory, the mass function is independent of the
power-spectrum. If the merging really occurs in an anisotropic way along
filaments, however, the filamentarity in the medium must affect the
final mass distribution of dark halos.  For example, for the case of a
power-law spectrum, the mass function might depend on the power-law
index sensitively, since the filamentarity of the density field should
depend on the power-law index.  We wish to present the effect of
anisotropic merging on the mass distribution of dark halos in the near
future (Lee \& Jing 2005, in preparation).

The discrepancy between our analytic model and the simulation results 
on the mass and redshift dependences implies that the shapes of dark halos 
are not fully determined by simply applying the Zel'dovich approximation. 
Rather, one has to take into account complicated nonlinear gravitational 
clustering like the anisotropic merging event. Nevertheless,
the overall agreement of our analytic model with the simulation results 
gives us a hope that our analytic model will be useful in quantifying how 
the initial cosmic web induces the ellipticity of dark matter halos, and 
provide a first theoretical step toward the goal of using the ellipticity 
distribution of galaxy clusters as a new cosmological probe 
(Lee 2005, in preparation).

\acknowledgments 
We thank an anonymous referee for very helpful suggestions.
J.L. and Y.P.J. thank Department of Physics at the University of Tokyo
for a warm hospitality where this work was initiated.  J.L.
acknowledges the research grant No. R01-2005-000-10610-0 from the Basic
Research Program of the Korea Science and Engineering
Foundation. Y.P.J. is partly supported by NKBRSF (G19990754), by NSFC
(Nos. 10125314,10373012), and by Shanghai Key Projects in Basic Research
(No. 04jc14079).  Work by Y.S. was supported in part by Grants-in-Aid
for Scientific Research from the Japan Society for Promotion of Science
(No.14102004, 16340053).

\clearpage
%%%%%%%%%%%%%%%%%%%%%%%%%%%%%%%%%%%%%%%%%%%%%%%%%%%%%%%%%%%%%%%%%%%
%%%%%%%%%%%%%%%%%%%%%%%%%%%%%%%%%%%%%%%%%%%%%%%%%%%%%%%%%%%%%%%%%%%

\clearpage
%%%%%%%%%%%%%%%%%%%%%%%%%%%%%%%%%%%%%%%%%%%%%%%%%%%%%%%%%%%%%%%%%%%
%%%%%%%%%%%%%%%%%%%%%%%%%%%%%%%%%%%%%%%%%%%%%%%%%%%%%%%%%%%%%%%%%%%

\clearpage
%%%%%%%%%%%%%%%%%%%%%%%%%%%%%%%%%%%%%%%%%%%%%%%%%%%%%%%%%%%%%%%%%%%
%%%%%%%%%%%%%%%%%%%%%%%%%%%%%%%%%%%%%%%%%%%%%%%%%%%%%%%%%%%%%%%%%%%

\clearpage
%%%%%%%%%%%%%%%%%%%%%%%%%%%%%%%%%%%%%%%%%%%%%%%%%%%%%%%%%%%%%%%%%%%
%%%%%%%%%%%%%%%%%%%%%%%%%%%%%%%%%%%%%%%%%%%%%%%%%%%%%%%%%%%%%%%%%%%

\clearpage
%%%%%%%%%%%%%%%%%%%%%%%%%%%%%%%%%%%%%%%%%%%%%%%%%%%%%%%%%%%%%%%%%%
%%%%%%%%%%%%%%%%%%%%%%%%%%%%%%%%%%%%%%%%%%%%%%%%%%%%%%%%%%%%%%%%%%%

\end{document}